\def\rfr#1{eq. (\ref{#1})}

\def\dert#1#2{\frac{{{d}}{#1}}{{{d}}{#2}}}              

\def\bar{\begin{eqnarray}}
\def\ear{\end{eqnarray}}
\def\bb{\bibitem}
\def\eqi{\begin{equation}}
\def\eqf{\end{equation}}
\def\eqia{\begin{eqnarray}}
\def\eqfa{\end{eqnarray}}
\def\rp#1#2{{#1\over#2}}

\def\lb#1{\label{#1}}

\def\psr{PSR J0737-3039A/B\ }

\def\cen{(1-e^2)}

\def\nkep{n^{(0)}}
\def\Pkep{P^{(0)}}
\def\Po{P_{\rm b}}
\def\Ts{T_{\odot}}
\def\xa{x_{\rm A}}
\def\xb{x_{\rm B}}
\def\Pop{\rp{\Po}{2\pi}}

\def\opn{\dot\omega_{\rm 1PN}}

\def\oc2{$\mathcal{O}(c^{-2})$}

\documentclass{elsart5p}
\usepackage{ifpdf}
\usepackage{graphicx,natbib,amssymb,lineno}
\ifpdf
\usepackage[%
  pdftitle={Instructions for use of the document class
    elsart},%
  pdfauthor={Simon Pepping},%
  pdfsubject={The preprint document class elsart},%
  pdfkeywords={instructions for use, elsart, document class},%
  pdfstartview=FitH,%
  bookmarks=true,%
  bookmarksopen=true,%
  breaklinks=true,%
  colorlinks=true,%
  linkcolor=blue,anchorcolor=blue,%
  citecolor=blue,filecolor=blue,%
  menucolor=blue,pagecolor=blue,%
  urlcolor=blue]{hyperref}
\else
\usepackage[%
  breaklinks=true,%
  colorlinks=true,%
  linkcolor=blue,anchorcolor=blue,%
  citecolor=blue,filecolor=blue,%
  menucolor=blue,pagecolor=blue,%
  urlcolor=blue]{hyperref}
\fi

\makeatletter
\def\elsartstyle{%
    \def\normalsize{\@setfontsize\normalsize\@xiipt{14.5}}
    \def\small{\@setfontsize\small\@xipt{13.6}}
    \let\footnotesize=\small
    \def\large{\@setfontsize\large\@xivpt{18}}
    \def\Large{\@setfontsize\Large\@xviipt{22}}
    \skip\@mpfootins = 18\p@ \@plus 2\p@
    \normalsize
}
\@ifundefined{square}{}{}
\makeatother

\journal{New Astronomy}
\begin{document}

\begin{frontmatter}
\title{Constraining the cosmological constant and the DGP gravity with the double pulsar PSR J0737-3039}

\author{Lorenzo Iorio\corauthref{boh}}
\address{Istituto Nazionale di Fisica Nucleare (INFN), Sezione di Pisa. Address for correspondence: Viale Unit$\grave{a}$ di Italia 68, 70125\\Bari (BA), Italy}

\corauth[boh]{Corresponding author}
\ead{lorenzo.iorio@libero.it}

\begin{abstract}
We consider the double pulsar \psr\ binary system as a laboratory to locally  test the orbital effects induced by  an uniform cosmological constant $\Lambda$ in the framework of the known general relativistic laws of gravity, and the DGP braneworld  model of gravity independently of the cosmological acceleration itself for which they were introduced.  We, first, construct the ratio $R=\Delta\dot\omega/\Delta P$ of the discrepancies between the phenomenologically determined  periastron rate $\dot\omega$  and orbital period $\Po$ and their predicted  values from  the 1PN $\opn$ approximation and the third Kepler law $\Pkep$. Then, we compare its value $|R| = (0.3\pm 4) \times 10^{-11} \ {\rm s}^{-2}$, compatible with zero within the errors, to the ratios $R_{\Lambda}$ and $R_{\rm DGP}$ of the effects induced on the apsidal rate and the orbital period by $\Lambda$ and the DGP gravity; we find them neatly incompatible with $R$ being $R_{\rm \Lambda} = (3.4\pm 0.3) \times 10^{-8} \ {\rm s}^{-2}$ and $R_{\rm DGP} = (1.4\pm 0.1) \times 10^{-7} \ {\rm s}^{-2}$, respectively. Such a result, which for the case of $\Lambda$ is valid also for any other Hooke-like exotic force proportional to $r$, is in agreement with other negative local tests recently performed in the Solar System with the ratios of the non-Newtonian/Einsteinian perihelion precessions for several pairs of planets.
 \end{abstract}

\begin{keyword}
 Experimental tests of gravitational theories; Dark energy; Modified theories of gravity;  Pulsars
\PACS 04.80.Cc, 95.36.+x, 04.50.Kd,  97.60.Gb
\end{keyword}
\end{frontmatter}

\section{Introduction}
Since, at present, the only reason why the cosmological constant\footnote{See \citep{Cal08} and references therein for an interesting historical overview.} $\Lambda$ is believed to be nonzero relies upon the observed acceleration of the universe \citep{Rie98,Per99}, i.e. just the phenomenon for which $\Lambda$ was introduced (again), it is important to find independent observational tests of the existence of such an exotic component of the spacetime.

In this paper we wish to put on the test the hypothesis that $\Lambda\neq 0$, where $\Lambda$ is the uniform cosmological constant of the Schwarzschild-de Sitter \citep{Stu99} (or Kottler \citep{Kot18}) spacetime, by suitably using the latest determinations of the parameters (see Table \ref{tavola}) of the double pulsar \psr\ system \citep{Bur03}.
The approach followed here consists in deriving analytical expressions $\mathcal{O}_{\Lambda}$ for the effects induced by $\Lambda$ on some quantities for which empirical values $\mathcal{O}_{\rm meas}$ determined  from fitting the timing data exist.  By taking into account the known classical and relativistic effects $\mathcal{O}_{\rm known}$ affecting such quantities, the discrepancy $\Delta{\mathcal{O}}=\mathcal{O}_{\rm meas} - \mathcal{O}_{\rm known}$ is constructed and attributed to the action of $\Lambda$, which was not modelled in the pulsar data processing. Having some $\Delta\mathcal{O}$ and $\mathcal{O}_{\Lambda}$ at hand, a suitable combination $\mathcal{C}$, valid just for the case $\Lambda\neq 0$, is constructed out of  them in order to  compare $\mathcal{C}_{\rm meas}$      to $\mathcal{C}_{\rm \Lambda}$: if the hypothesis $\Lambda\neq 0$ is correct, they must be equal within the errors. Here we will use the anomalistic period $\Po$ and the periastron precession $\dot\omega$ for which purely phenomenological determinations exist in such a way that our $\mathcal{C}$ is the ratio of $\Delta\dot\omega$ to $\Delta\Po$; as we will see, this observable is independent of $\Lambda$ but, at the same time, it retains a functional dependence on the system's parameters peculiar to the $\Lambda-$induced force and of any other Hooke-like forces.

The present  work complements \citep{Ior07} in which a similar test was conducted in the Solar System by means of the latest determinations of the
secular precessions of the longitudes of the perihelia of several planets. The result of \citep{Ior07} was negative for the Schwarzschild-de Sitter spacetime with uniform $\Lambda$; as we will see, the same conclusion will be traced out of this paper in Section \ref{posec}.

A complementary approach to explain the cosmic acceleration without resorting to dark energy was followed by Dvali, Gabadadze and Porrati (DGP) in their braneworld modified model of gravity \citep{DGP}.   Among other things, it predicts effects which could be tested on a local, astronomical  scale.
In \citep{Ior07} a negative test in the Solar System was reported; as we will see in Section \ref{dgpsec},  \psr\ confirms such a negative outcome at a much more stringent level.

The conclusions are in Section \ref{conc}.

\section{The effect of $\Lambda$ on the periastron and the orbital period}
The Schwarzschild-de Sitter metric induces an extra-acceleration\footnote{The present test is valid for all exotic Hooke-type forces of the form $Cr$ \citep{Cal08}, with $C$ arbitrary nonzero constant.}
\citep{Rin} \eqi \mathbf{A}_{\Lambda}=\rp{1}{3}\Lambda c^2 \mathbf{r},\lb{acc}\eqf
where $c$ is the speed of light; \rfr{acc}, in view of the extreme smallness of the assumed nonzero value cosmological constant ($\Lambda \approx 10^{-52}$ m$^{-2}$), can be treated perturbatively with the standard techniques of celestial mechanics.  In \citep{Hau03} the secular precession of the pericentre of  a test particle around a central body of mass $\mathfrak{M}$ was found    to be
\eqi\dot\omega_{\Lambda} = \rp{\Lambda c^2}{2 \nkep}\sqrt{1-e^2},\lb{olam}\eqf where
\eqi\nkep = \sqrt{\rp{G\mathfrak{M}}{a^3}}\eqf  is the Keplerian mean motion; $a$ and $e$ are the semimajor axis and the eccentrity, respectively, of the test particle's orbit.
Concerning a binary system, in \citep{Ser06} it was shown that the equations for the relative motion are those of a test particle in a Schwarzschild-de Sitter space-time with a
source mass equal to the total mass of the two-body system, i.e. $\mathfrak{M} = \mathfrak{m}_{\rm A}+\mathfrak{m}_{\rm B}$. Thus, \rfr{olam} is valid in our case; $a$ is the semi-major axis of the
relative orbit.

Following the approach by \citet{Ser06}, we will now compute $P_{\Lambda}$, i.e. the contribution of $\Lambda$ to the orbital period.
One of the six Keplerian orbital elements in terms of which it is
possible to parameterize the orbital motion in a
binary system is the mean anomaly  $\mathcal{M}$ defined as
$\mathcal{M}\equiv n(t-T_0)$, where $n$ is the mean motion and
$T_0$ is the time of pericenter passage. The mean motion $n\equiv
2\pi/ P_{\rm b}$ is inversely proportional to the time elapsed
between two consecutive crossings of the pericenter, i.e. the
anomalistic period $P_{\rm b}$. In Newtonian mechanics, for two
point-like bodies, $n$ reduces to the usual Keplerian expression
$\nkep=2\pi/\Pkep$. In
many binary systems, as in the double pulsar one, the period $P_{\rm b }$ is  accurately
determined in a phenomenological, model-independent way, so that, in principle,
it accounts for all the dynamical features of the system, not only
those coming from the Newtonian point-like terms, within the
measurement precision.

The Gauss equation for the variation of the mean anomaly, in the case of an entirely radial disturbing acceleration   $A$
like \rfr{acc}, is
\eqi\dert{\mathcal{M}} t=n-\rp{2}{na}A
\left(\rp{r}{a}\right)+\rp{(1-e^2)}{nae}A\cos f,\lb{gauss}\eqf
where $f$ is the true anomaly, reckoned from the pericenter.
Using the eccentric anomaly $E$, defined as
\eqi \mathcal{M} = E - e\sin E,\eqf turns out to be more convenient.
The unperturbed Keplerian ellipse, on which the right-hand-side of \rfr{gauss} must be evaluated, is
\eqi r = a\left(1-e\cos E\right);\eqf by using
\rfr{acc} and \begin{equation}\left\{\begin{array}{lll}
\dert{\mathcal{M}}E = 1 - e\cos E,\\\\
\cos f = \rp{\cos E - e}{1 - e\cos E},
\lb{grazia}
\end{array}\right.\end{equation}
\rfr{gauss} becomes
\begin{eqnarray} & \dert E t & =  \rp{\nkep}{\left(1-e\cos E\right)}\left\{1-\rp{\Lambda c^2}{3{\nkep}^2}\left[2\left(1-e\cos E\right)^2 -\right.\right.\nonumber\\
& &-\left.\left.\rp{\left(1-e^2\right)}{e}\left(\cos E-e\right)\right]\right\}.\lb{grossa}
\end{eqnarray}
Since $\Lambda c^2/3{\nkep}^2\approx 10^{-29}$ from \rfr{grossa} it can be obtained
\begin{eqnarray} & \Po & \simeq \int_0^{2\pi}\rp{\left(1-e\cos E\right)}{\nkep}\left\{1+\rp{\Lambda c^2}{3{\nkep}^2}\left[2\left(1-e\cos E\right)^2 -\right.\right.\nonumber\\
& &-\left.\left. \rp{\left(1-e^2\right)}{e}\left(\cos E-e\right)\right]\right\}dE,
\end{eqnarray}
which integrated yields   that
\eqi\Po = \Pkep + P_{\Lambda}\eqf
with
\eqi P_{\Lambda} = \rp{\pi\Lambda c^2\left(7 + 3 e^2\right)}{3{\nkep}^3}.\lb{pla}\eqf

\subsection{Combining the periastron and the orbital period}\lb{posec}
The general relativistic expressions of the post-Keplerian parameters $r,s$ and $\dot\omega$ are
 \begin{equation}\left\{\begin{array}{lll}
r=T_{\odot}m_{\rm B},\\\\
s = x_{\rm A}\left(\rp{P_{\rm b}}{2\pi}\right)^{-2/3} T_{\odot}^{-1/3}M^{2/3}m_{\rm B}^{-1}, \\\\
\opn = \rp{3}{\cen}\left(\Pop\right)^{-5/3}(\Ts M)^{2/3},
\lb{shap}
\end{array}\right.\end{equation}
where \eqi \Ts = \rp{G{\mathfrak{M}}_{\odot}}{c^3}\eqf and $M=m_{\rm A} + m_{\rm B}$ in units of solar masses.

By means of \eqi a = \rp{c}{s}(\xa + \xb)\eqf and of the equations for $r$ and $s$ it is possible to express $\Pkep$ and $\opn$ in terms of $\Po$ and of the phenomenologically determined Keplerian and post-Keplerian parameters $\xa, \xb, r, s$ as
\begin{equation}\left\{\begin{array}{lll}
\Pkep = 2\left(\rp{2}{\Po}\right)^{1/2}\left[\pi(\xa+\xb)\right]^{3/2}\left(\rp{\xa}{r}\right)^{3/4}s^{-9/4},\\\\
\opn = \rp{3sr}{\xa\cen}\left(\Pop\right)^{-1}.
\lb{grazia}
\end{array}\right.\end{equation}

 In such a way we can genuinely compare them to $\Po$ and $\dot\omega$ because they do not contain quantities obtained from the third Kepler law and the general relativistic periastron precession themselves; moreover, we have expressed the sum of the masses entering both $\Pkep$ and $\opn$ in terms of $r$ and $s$, thus avoiding any possible reciprocal imprinting between the third Kepler law and the periastron rate.
At this point it is possible to construct
\eqi R \equiv \rp{\Delta\dot\omega}{\Delta P},\lb{rap}\eqf with
 \begin{equation}\left\{\begin{array}{lll}
 \Delta\dot\omega = \dot\omega-\opn,\\\\
\Delta P = \Po-\Pkep;
 \lb{rapp}
\end{array}\right.\end{equation}
note that \eqi R = R(\Po,\xa,\xb,e;\dot\omega,r,s).\eqf
By attributing   $\Delta\dot\omega$ and $\Delta P$ to the action of $\Lambda$, not modelled into the routines used to fit the \psr\ timing data, it is possible to compare $R$ to
\eqi R_{\Lambda}\equiv\rp{\dot\omega_{\Lambda}}{P_{\Lambda}}=\rp{3\sqrt{1-e^2}\Po r^{3/2}s^{9/2}}{4\pi^2(7 + 3e^2)(\xa+\xb)^3 \xa^{3/2}}\lb{Rlam}\eqf
and see if \rfr{rap} and \rfr{Rlam} are equal within the errors. Note that \rfr{Rlam} is independent of $\Lambda$ and, by definition, is able to test the hypothesis that $\Lambda\neq 0$. From Table \ref{tavola} it turns out
\eqi R_{\Lambda}=(3.4\pm 0.3)\times 10^{-8}\ {\rm s}^{-2};\eqf $R_{\Lambda}$ is a well determined quantity, different from zero at about 11 sigma level.
In regard to $R$ we have
 \begin{equation}\left\{\begin{array}{lll}
\Delta\dot\omega = -0.3\pm 2.1\ {\rm deg\ yr}^{-1},\\\\
\Delta P = 59\pm 364\ {\rm s},
\lb{rappnum}
\end{array}\right.\end{equation}
so that \eqi \left|R\right| = (0.3\pm 4)\times 10^{-11}\ {\rm s}^{-2}; \eqf $R$ is compatible with zero in such a way that its range does not overlap with the one of $R_{\Lambda}$: indeed, the upper bound on $R$ is three orders of magnitude smaller than the lower bound on $R_{\Lambda}$.
Thus, we must conclude that\footnote{In principle, also the 1PN correction to the third Kepler law \citep{Dam86} should be included in $\Delta P$, but it does not change the result.}
\eqi R \neq R_{\Lambda}.\eqf   Concerning the released uncertainties in $R$ and $R_{\Lambda}$, they must be considered as upper bounds since they have been conservatively computed by linearly adding the individual biased terms due to $\delta\Po,\delta\dot\omega,\delta e, \delta\xa,\delta\xb,\delta r,\delta s$ in order to account for the existing correlations \citep{Kra06} among  them.

The results of the present study confirm those obtained in the Solar System by taking the ratio of the estimated corrections to the standard Newtonian/Einsteinian precessions of the longitude of the perihelia $\varpi$ for different pairs of planets \citep{Ior07}.    It would be very interesting to devise analogous tests involving other observables (lensing, time delay) affected by $\Lambda$ as well recently computed in, e.g., \citep{Rug07, Ser08}.
\section{The Dvali-Gabadadze-Porrati braneworld model}\lb{dgpsec}
The approach previously outlined for $\Lambda$ can be followed also for the DGP braneworld model \citep{DGP} which recently received great attention from an observational point of view \citep{Dva03,Ior08}.

The preliminary confrontations with data so far performed refer to the perihelia of the Solar System planets. Indeed, DGP gravity predicts an extra-precession of the pericentre of a test particle   \citep{Lue03,Ior05}
\eqi \dot\omega_{\rm DGP}=\mp\rp{3}{8}\left(\rp{c}{r_0}\right)\left(1-\rp{13}{32}e^2\right),\lb{odgp}\eqf where the signs $\mp$ are related to the two different cosmological branches of the model and $r_0$ is a free-parameter set to about 5 Gpc by Type IA Supernov{\ae} data, independent of the orbit's semimajor axis. The predicted precessions of about $10^{-4}$ arcsec cy$^{-1}$ were found to be compatible with the estimated  corrections to the usual apsidal precessions of planets considered one at a time separately \citep{Ior08}, but marginally incompatible with the ratio of them for some pairs of inner planets \citep{IorAHEP}.

The effects of DGP model on the orbital period is    \citep{Ior06}
\eqi P_{\rm DGP}=\mp\rp{11}{8}\pi\left(\rp{c}{r_0}\right)\rp{a^3 (1-e^2)^2}{G\mathfrak{M}}\lb{pdgp}.\eqf

From \rfr{odgp} and \rfr{pdgp} it is possible to construct
\eqi R_{\rm DGP}\equiv\rp{\dot\omega_{\rm DGP}}{P_{\rm DGP}},\eqf which, expressed in terms of the phenomenologically determined parameters of \psr, becomes
\eqi R_{\rm DGP} = \rp{3\left(1-\rp{13}{32}e^2\right)\Po r^{3/2} s^{9/2}}{22\pi (1-e^2)\left(\xa + \xb\right)^3 \xa^{3/2}}.\lb{Rdgp}\eqf
Putting the figures of Table \rfr{tavola} into \rfr{Rdgp} and computing the uncertainty as done in the case of $\Lambda$ yields
\eqi R_{\rm DGP} =  (1.4\pm 0.1)\times 10^{-7}\ {\rm s}^{-2}.\eqf
As can be noted, the lower bound of $R_{\rm DGP}$ is four orders of magnitude larger than the upper bound of $R$, so that we must conclude that, also in this case,
\eqi R\neq R_{\rm DGP}.\eqf
The outcome by \citet{IorAHEP} is, thus, confirmed at a much more stringent level. 

An analysis of type Ia supernov{\ae} (SNe Ia) data disfavoring  DGP model can be found in \citep{Ben05}.
\section{Conclusions}\lb{conc}
In this paper we used the most recent determinations of the orbital parameters of the double pulsar binary system \psr\ to perform local tests of  two complementary approaches to the issue of the observed acceleration of the universe: the uniform cosmological constant $\Lambda$ in the framework of the known general relativistic laws of gravity  and the multidimensional braneworld model by Dvali, Gabadadze and Porrati which, instead, resorts to a modification of the currently known laws of gravity. Since, at present, there are no observational evidences for such theoretical schemes other than just the cosmological phenomenon for which they were introduced, it is important to put them on the test independently of the cosmological acceleration itself.
It is worthwhile noting that the results for $\Lambda$ hold also for any other Hooke-like additional force proportional to $r$.

To this aim, we  considered the phenomenologically determined  the periastron precession $\dot\omega$ and the orbital period $\Po$ of \psr\ by contrasting them to the predicted  1PN periastron rate $\opn$ and the Keplerian period $\Pkep$. With such discrepancies we constructed the ratio $R=\Delta\omega/\Delta P$  by finding it compatible with zero: $|R| = (0.3\pm 4) \times 10^{-11} \ {\rm s}^{-2}$.  Then, we compared $R$ to the predicted ratios for the effects of $\Lambda$ and the DGP gravity-not modeled in the pulsar data processing-on the periastron rate and the orbital period by finding $R_{\Lambda} = (3.4\pm 0.3) \times 10^{-8} \ {\rm s}^{-2}$ and $R_{\rm DGP} = (1.4\pm 0.1) \times 10^{-7} \ {\rm s}^{-2}$, respectively. Thus, the outcome of such a local test is neatly negative, in agreement with other local tests recently performed in the Solar System by taking the ratio of the non-Newtonian/Einsteinian rates of the perihelia for several pairs of planets.

 \section*{Acknowledgments}
I thank O. Bertolami for useful references.

\newpage

\begin{table}
 \caption{ Relevant Keplerian and post-Keplerian parameters of the binary system
\psr \citep{Kra06}. The orbital
period $P_{\rm b}$ is measured with a precision of $4\times 10^{-6}$ s. The projected semimajor axis is
defined as $x=(a_{\rm bc}/c)\sin i,$  where $a_{\rm bc}$ is the
barycentric semimajor axis, $c$ is the speed of light and $i$ is the angle
between the plane of the sky, perpendicular to the line-of-sight,
and the orbital plane. The eccentricity is $e$. The best determined post-Keplerian parameter is, to date, the periastron rate $\dot\omega$ of the component A. The phenomenologically determined post-Keplerian parameter $s$, related to the general relativistic Shapiro time delay,
is equal to $\sin i$; we  have
conservatively quoted the largest error in $s$ reported in
\citep{Kra06}.  The other post-Keplerian parameter related to the Shapiro delay, which is used in the text, is $r$.
}
\label{tavola}

\begin{tabular}{lllllll}

\noalign{\hrule height 1.5pt}

$P_{\rm b}$ (d)& $x_{\rm A}$ (s) & $x_{\rm B}$ (s) & $e$ & $\dot\omega$ (deg yr$^{-1}$) & $s$ & $r$ ($\mu$s)\\
\hline
$0.10225156248(5)$ & $1.415032(1)$ & $1.5161(16)$ &  $0.0877775(9)$ & $16.89947(68)$  & $0.99974(39)$  &  $6.21(33)$ \\
\hline

\noalign{\hrule height 1.5pt}

\end{tabular}

\end{table}

\end{document}